\documentclass[proof]{WileyASNA-v1}

\articletype{Article Type}%

\received{26 April 2016}
\revised{6 June 2016}
\accepted{6 June 2016}

\begin{document}

\title{The broad emission line asymmetry in low mass ratio
of supermassive binary black holes on elliptical orbits}

\author[1]{Sa\v sa Simi\'c*}

\author[2]{Luka \v C. Popovi\'c}

\author[3]{Andjelka Kova\v cevi\'c}

\author[3]{Dragana Ili\'c}

\address[1]{\orgdiv{Faculty of science}, \orgname{University of Kragujevac}, \orgaddress{\state{Radoja Domanovi\'ca 12, 34000 Kragujevac}, \country{Serbia}}}

\address[2]{\orgdiv{Astronomical observatory}, \orgname{University of Belgrade}, \orgaddress{\state{Volgina 7, 11000 Belgrade}, \country{Serbia}}}

\address[3]{\orgdiv{Faculty of mathematics}, \orgname{University of Belgrade}, \orgaddress{\state{Studentski trg 16}, \country{Serbia}}}

\corres{*Sa\v sa Simi\'c, \email{ssimic@uni.kg.ac.rs}}


\abstract{We investigate the broad line profiles emitted from a system supermassive binary black hole (SMBBH) having elliptical orbits and low mass ratio of $m_2/m_1\sim 0.1$. Our model assumes
a super Eddington accretion flow in the case of a smaller component, whereas the massive component has very small or negligible accretion, therefore supposing that no broad line region (BLR) is attached to it. Thus, the proposed SMBBH system contains one moving BLR, associated with the less massive component and one circum-binary BLR. We study the effect of different total mass of the system (ranging from 10$^6$ to 10$^8$ Solar masses) to the $\mathrm{H\beta}$ line profiles and to the continuum and line light curves. The resulted broad line profiles are asymmetric and shifted, and are varying during the orbital period. The asymmetry in the broad line profiles is discussed in terms of expected differences between the proposed model of the SMBBH with one active component and the scenario of a recoiling black hole. We discuss the periodicity detected in the line and continuum light curves, as well as in the variations of the line asymmetry and shift.}


\maketitle

\footnotetext{\textbf{Abbreviations:} AGN, active galactic nucleus; BLR, broad line region; cBLR, circumbinary Broad Line Region; GW, gravitational wave; SMBH, supermassive blak hole; SMBBH, supermassive binary black hole;   SPD, Spectral Power Distribution; LS, Lomb Scargle; }

\section{Introduction}
\label{sec:Intro}

The last stage of galaxy collisions probably results in formation of a sub-pc super-massive binary black hole (SMBBH) system \citep{beg80}, which may be ending with the coalescence of two super-massive black holes (SMBHs) and emission of low-frequency gravitational waves (GWs). As theoretical candidates for low-frequency GWs, the sub-pc SMBBHs may be important targets for GW detection facilities,  such as the Pulsar Timing Array project \citep[see e.g.,][]{nguyen20a} or Lunar Gravitational-Wave Antenna \citep[see e.g.,][]{Harms21}.

In the near coalescence phase, the sub-pc SMBBHs surrounded by gas may show an activity that is similar to active galactic nucleus (AGN), emitting specific and variable optical/UV continuum and line profiles \citep[see e.g.,][]{gaskel83,pop00,pop12,simic16,pop21,smail15}.
This specific variability can be used to detect an SMBBH presence \citep[see e.g.,][]{bon12, run15, barth15, shap16, li16, kov17, kov19, kov20, pop21}. After the coalescence, the newly formed SMBH  could be kicked from the host galaxy center with high velocity \citep[see e.g.,][]{guedes11},
and travel through a gas-rich region emitting a specific spectrum with peculiar broad line profiles. Several cases in which the AGN has an off-center location in the host galaxy have been detected, one being the quasar E1821+643 that shows asymmetric and redshifted broad lines \citep[see][]{robinson10,shapovalova16,jadhav21}.
However, the high broad line asymmetry may also emerge from a sub-pc SMBBH. in which only one component has a broad line region \citep[BLR, see Fig. 3 in][]{pop12}. Besides the complex broad emission line profiles, the sub-pc SMBBH system
may exhibit other specific characteristics  caused by dynamical effects, as e.g. periodicity \citep[for more details see][and reference therein]{pop21}.

In this paper, we explore the variability of sub-pc SMBBH system using the theoretical model presented in \cite{pop21}. In contrary to that model, here we consider a large difference in the SMBBH component masses ($m_1, m_2$), taking that their mass ratio $q=m_2/m_1\sim 0.1$. This scenario could result in a smaller SMBH "cleaning" the gas around the more massive SMBH, thus leading to the massive SMBH having lower accretion rate and not forming its own BLR, whereas the smaller SMBH should have a BLR \cite[see e.g.,][]{mils01,khan13,vasil15}. We assume that the smaller SMBH component is in the super Eddington regime, with accretion rate much higher than in the more massive component. We simulate the evolution of the binary system during a couple of full orbits, and discuss the resulting observables, such as the mean and \textit{rms} H$\beta$ line profiles, the variations in the H$\beta$ line asymmetry and shift, as well as the line and continuum flux variability.

The paper is organized as following: in \S 2 we describe the model, in \S 3 we give the results of modeling and in \S 4 we outline our conclusions.

\section{Model}
\label{sec:Model}

The model is based on our earlier investigation of spectral variability of sub-pc SMBBHs due to their interaction and motion  \citep[in more details described in][hereafter referred to as PoSKI model]{pop21}.
In brief, the PoSKI model assumes a system of two SMBHs at sub-pc distance \citep[see Fig. 1 in][]{pop21}, in which each component has an accretion disc emitting broad-band continuum.
Each disc continuum emission is ionizing the surrounding gas forming local BLRs (i.e. BLR1,2), but also the total disc continuum emission is ionizing the gas around the system forming a circum-binary BLR (cBLR). Such a scenario can produce broad emission lines composed from three emitting components: the emission from the BLR1 and BLR2, which are moving  along with the two SMBHs, and the emission coming from a stationary cBLR \citep[see for more details][]{pop21}.

A special case of the proposed model is the SMBBH system with the large difference in the masses of components. Having the primary component ($m_1$) mass ten times larger than the mass of the secondary ($m_2$), one can expect the following scenario: a smaller component orbiting around the primary is wiping its surrounding gas. This may increase the disc accretion rate of the smaller component and produce stronger continuum emission, leading to the creation of the BLR2. The primary component  also has an accretion disc, but much smaller gas reservoir, which may cause  the decrease of the disc accretion rate and continuum emission, so that the BLR1 cannot be formed, i.e. a portion of the  total broad emission line, coming from the primary component, is missing. There are numerous studies of binary accretion disc configuration presented in the literature \citep[see e.g.,][]{artym94,farris14,miranda15,dorozo16,heath20} showing that the SMBH binary system accretion discs can have complex form, with number of different features, such as cavities or gaps in case of small mass ratio \citep{dorozo16}.

The PoSKI model \citep{pop21} belongs to
a group of models \citep[se e.q.][]{dorazio15,nguyen19,nguyen20a,nguyen20b} that predicts
periodicities in the emission line/continuum light curves and broad line parameters of SMBBHs.
There are significant similarities between the PoSKI and models implemented in mentioned papers. For example, our model uses the relativistic beaming effect proposed by \cite{dorazio15}, but in addition include three BLRs, of which two are connected with the moving SMBHs and one is circumbinary and stationary \citep[like in the ][]{nguyen16,nguyen19,nguyen20a,nguyen20b}.
In addition, in difference with these models, in the PoSKI model we consider that the size of the moving BLRs around each SMBH are defined by their Roche lobes (affected by relativistic beaming) and that the stationary cBLR dimension depends on the total disc continuum luminosity of the system. Moreover, the PoSKI model includes influence of the component masses and continuum luminosities to the broad line parameters taking empirical relations between the BLR radius, continuumm luminosity and SMBH mass \citep[similar as it is given in ][]{simic16}.

\begin{figure*}[ht!]
	\centerline{\includegraphics[width=12cm]{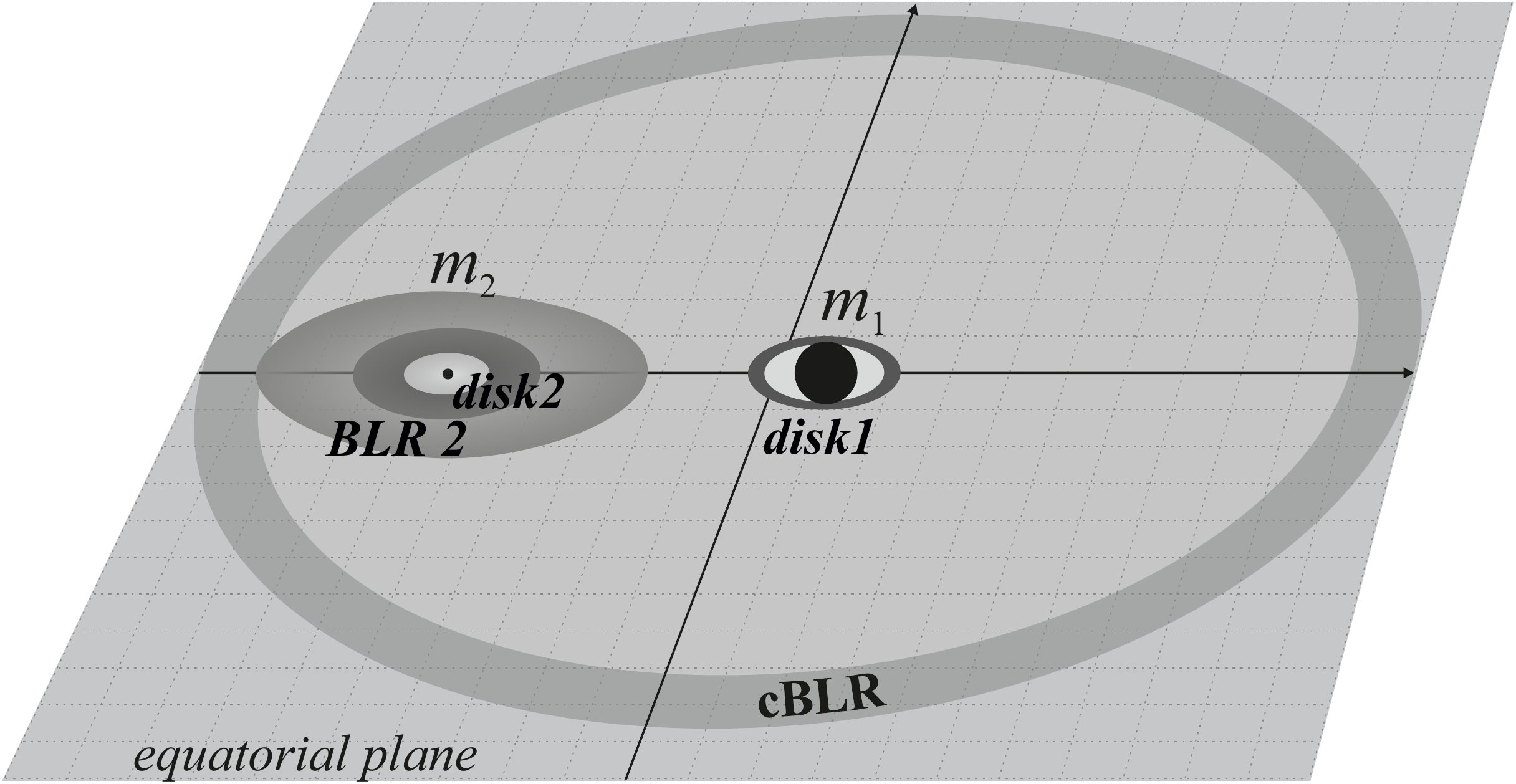}}
	\caption{Sketch of the SMBH binary system configuration used in this paper, with the primary, more massive component $m_1$ and a smaller secondary $m_2$. Orbits of SMBHs define the equatorial plane, and two orthogonal lines present local coordinates with the origin in the binary system baricenter. Corresponding accretion discs are denoted with \textit{disk1} and \textit{disk2}, and the BLR of the smaller component and circumbinary region with BLR2 and cBLR, respectively.}
	\label{fig:SMBBH_sketch}
\end{figure*}

\begin{table*}
\centering
\caption{Model input parameters used for calculations. The mass ratio is kept constant $q=m_2/m_1=0.1$, whereas the order of total mass magnitude, denoted with the degree index, is in the range of $x\in \{6,7,8\}$. Masses for SMBHs components are given as $m_1=10\times10^x\mathrm{M_{\odot}}$ and $m_2=1\times10^x\mathrm{M_{\odot}}$.}
\begin{tabular}{c|c|c|c|c|c|c}
$P_{x=6}[years]$ & $P_{x=7}[years]$ & $P_{x=8}[years]$ & R[pc] & $e_{cc}$ & $\theta[^o$] & $\omega[^o$] \\
\hline
 27.4 & 8.6 & 2.7 & 0.01 & 0.5 & 45 & 30 \\
\end{tabular}
\label{tab:inp_params}
\end{table*}

In Fig. \ref{fig:SMBBH_sketch} we present the scheme of the proposed SMBBH system that has a massive SMBH $m_1$ located close to the system barycenter  and one smaller component SMBH $m_2$ at some distance from the barycenter. In such a configuration, the bigger component is slowly wobbling around the barycenter, while the smaller component is attracting  and collecting the surrounding material creating its own BLR.  The disc continuum emission from both accretion discs is responsible for the creation of the cBLR that is surrounding both components.  In addition, due to the small amount of gas around the primary, its disc accretion rate may be smaller than in the secondary. Thus, in difference to the PoSKI model, the accretion rate coefficients \citep[Eq. B3, Appendix B in][]{pop21} of the components are set to be  different: the smaller component is assumed to be in the super Eddington regime, whereas the massive component has very low accretion rate (much below 0.3). All assumptions about the interaction between the components and dynamical effects are the same as in the PoSKI model, with the same co-planar geometry and both components orbiting in equatorial plane (see Fig. \ref{fig:SMBBH_sketch}). The accretion disks are denoted with \textit{disk1} and \textit{disk2}, the BLR of the smaller component and circumbinary region with BLR2 and cBLR, respectively (Fig. \ref{fig:SMBBH_sketch}). As in the PoSKI model, in order to have more realistic simulations, our simulations include the white noise, which is superposed to the simulated spectra on the level of 3\% of the maximal light curve variability.

\section{Results}
\label{sec:Results}

\begin{figure*}[h]
	\centerline{\includegraphics[width=16.5cm]{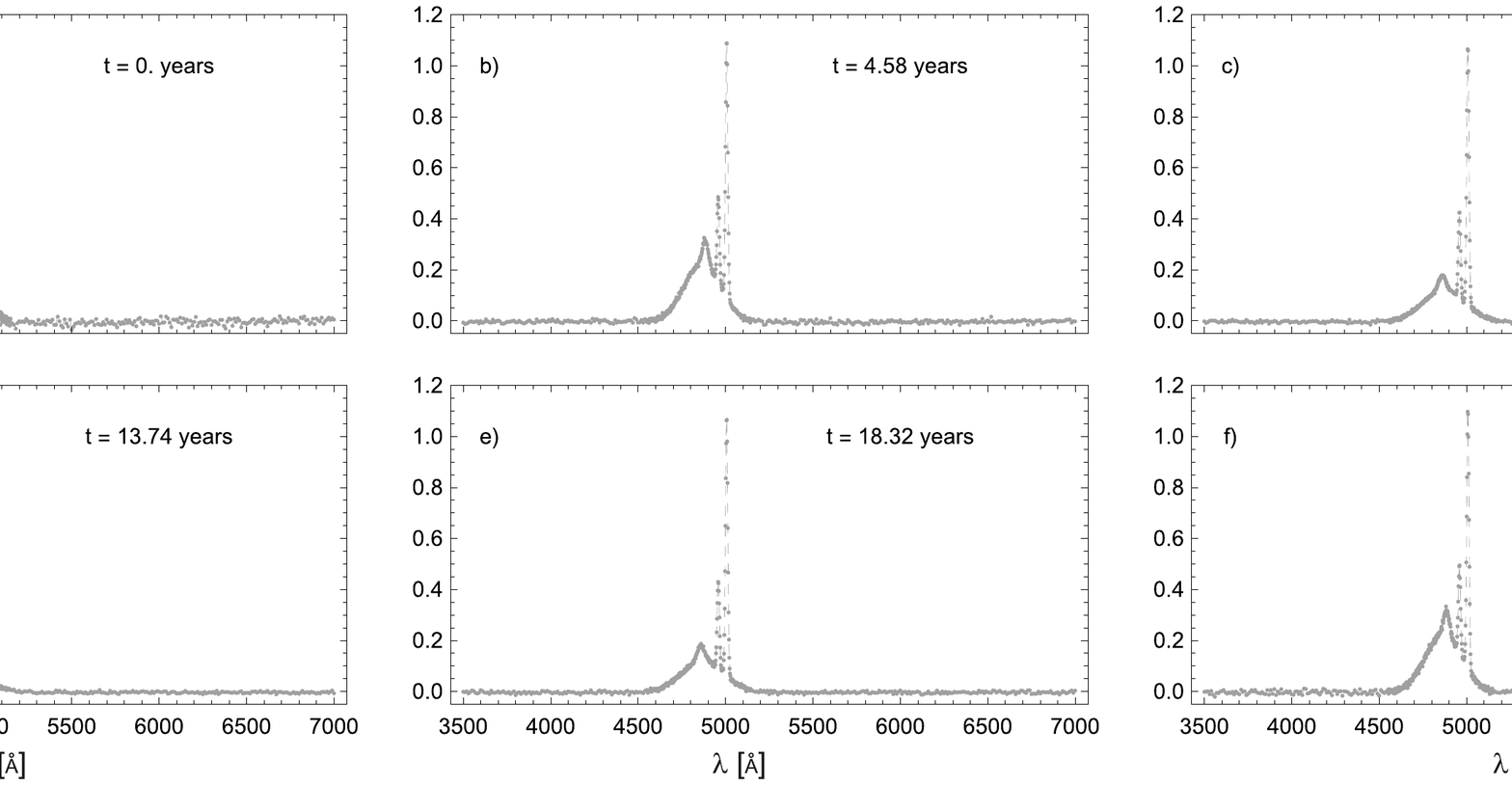}}
	\caption{H$\beta$ line profile emitted from six different positions during one full rotation of the SMBBH system with a total mass of $\sim 10^{6}\mathrm{M_{\odot}}$. Initial phase was taken when binaries are in an aphelion position, and time elapsed is indicated in the upper corner of each plot.}
	\label{fig:composite}
\end{figure*}

In order to explore spectral variability of the SMBBH system described above, we modeled the emission of H$\beta$ and continuum at 5100\AA \, during three full orbits of the SMBBH system. These spectral features were selected since they are usually considered in the AGN monitoring campaigns.
Fig. \ref{fig:composite} shows the modeled H$\beta$ line for six different phases during one full orbit of the system with the total mass of components of  $\sim10^6\mathrm{M_{\odot}}$. Initial phase was taken when binaries are in an aphelion position. To better illustrate the change in the broad line profile and intensity, we artificially added a constant emission from a narrow line region,  which  is typically illustrated with constant [OIII]$\lambda\lambda$4959,5007\AA\AA \, lines (see Fig. \ref{fig:composite}).

\begin{figure}[h]
	\centerline{\includegraphics[width=8.2cm]{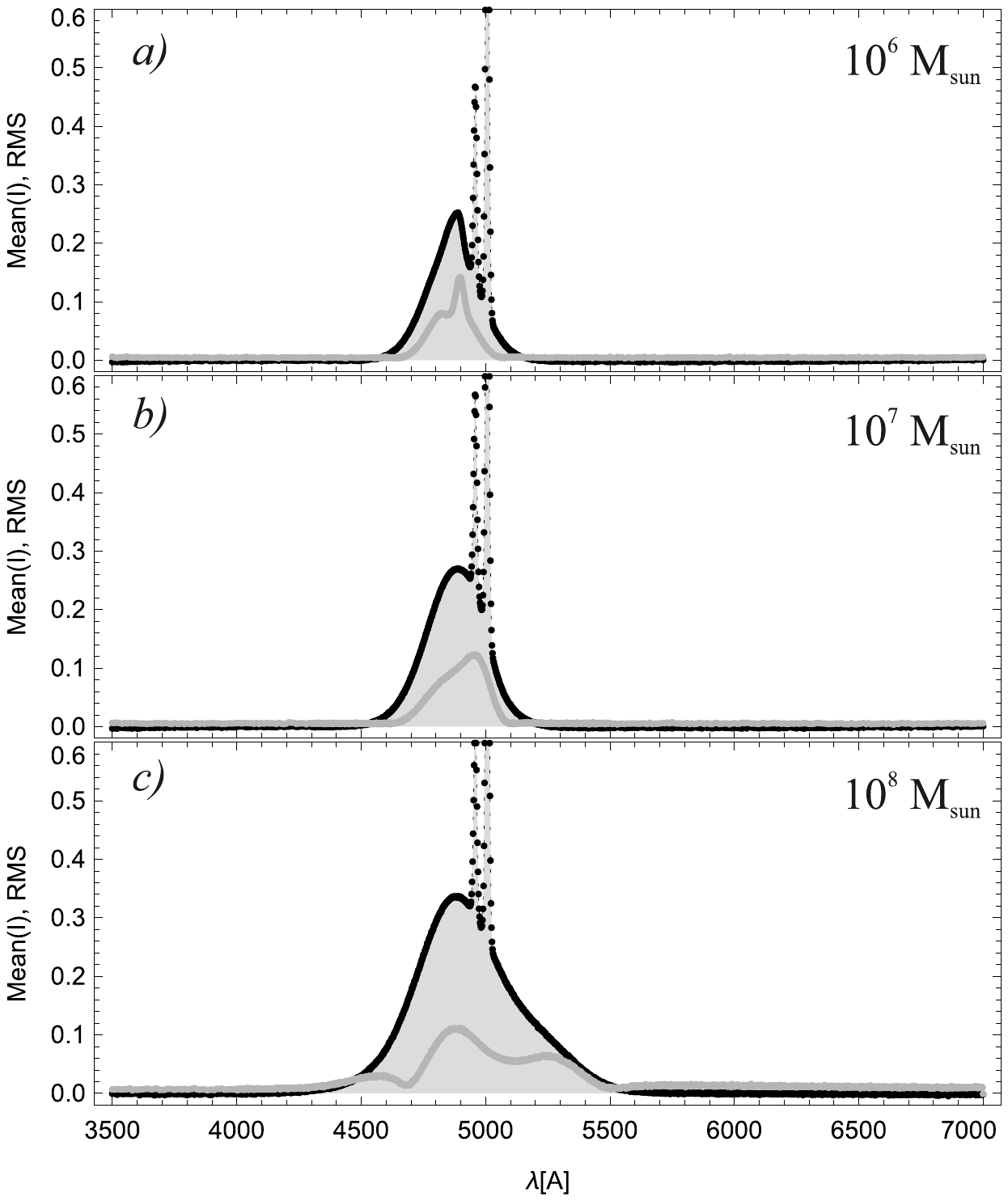}}
	\caption{Mean (black line) and corresponding \textit{rms} (gray line)  profile of the H$\beta$ line during three full orbits and for three different total masses of the SMBBH system:
	a) $10^{6}\mathrm{M_{\odot}}$, b) $10^{7}\mathrm{M_{\odot}}$ and c) $10^{8}\mathrm{M_{\odot}}$.}
	\label{fig:mr}
\end{figure}

Fig. \ref{fig:composite} shows that the line intensity and   profiles are changing during the period of a couple of years. The line emission is brighter when distance between the components is maximal, and the emission line remains weaker when components come closer. This is a result of to the components interaction, which is discussed in details the PoSKI model paper \citep{pop21}, but here is even more prominent due to the higher mass ratio of the components. Additionally, in this particular case with the SMBHs of the order of mass equal to $10^{6}\mathrm{M_{\odot}}$, the broad line component coming from the cBLR is much broader than the line component of the BLR of the secondary (BLR2).
This is caused by the assumption of the PoSKI model that the gas velocity in the BLR2 (which size is defined by the Roche lobe) depends on the second component mass, and the gas kinematics within the cBLR depends on the total mass of the binary system. Therefore, in this special case of low-mass secondary component, the gas within the cBLR is moving faster than in the BLR2, which results with the broader line emitted from the cBLR than from the BLR2. As a consequence, a superposed narrower spike arising from the BLR2 is clearly seen in the total line profile, and is shifting along the profile depending on the secondary SMBH orbital motion.

The H$\beta$ line flux is scaled to the narrow $\mathrm{OIII}$ doublet mean flux, which is assumed to be constant as it is originating from the narrow line region. The mean $\mathrm{OIII}$ line flux is calculated using the mean total continuum flux from the aphelion and perihelion position.

To explore the line profile variability we calculated the mean H$\beta$ line profile and the corresponding \textit{rms}. Fig. \ref{fig:mr} shows the mean and {\it rms} H$\beta$ line profiles  for three values of the total system mass.
The mean line profile becomes broader and more asymmetric for larger total mass of the system. For systems with the total mass of $\sim 10^6M_{\odot}$ the width of the line is significantly narrower than in the  system with the total mass of $\sim 10^8M_{\odot}$ (see Fig. \ref{fig:mr}). This is expected since in our model we assume that BLRs are virialized  \citep[see][]{pop21}.

\begin{figure*}[h]
	\centerline{\includegraphics[width=16.4cm]{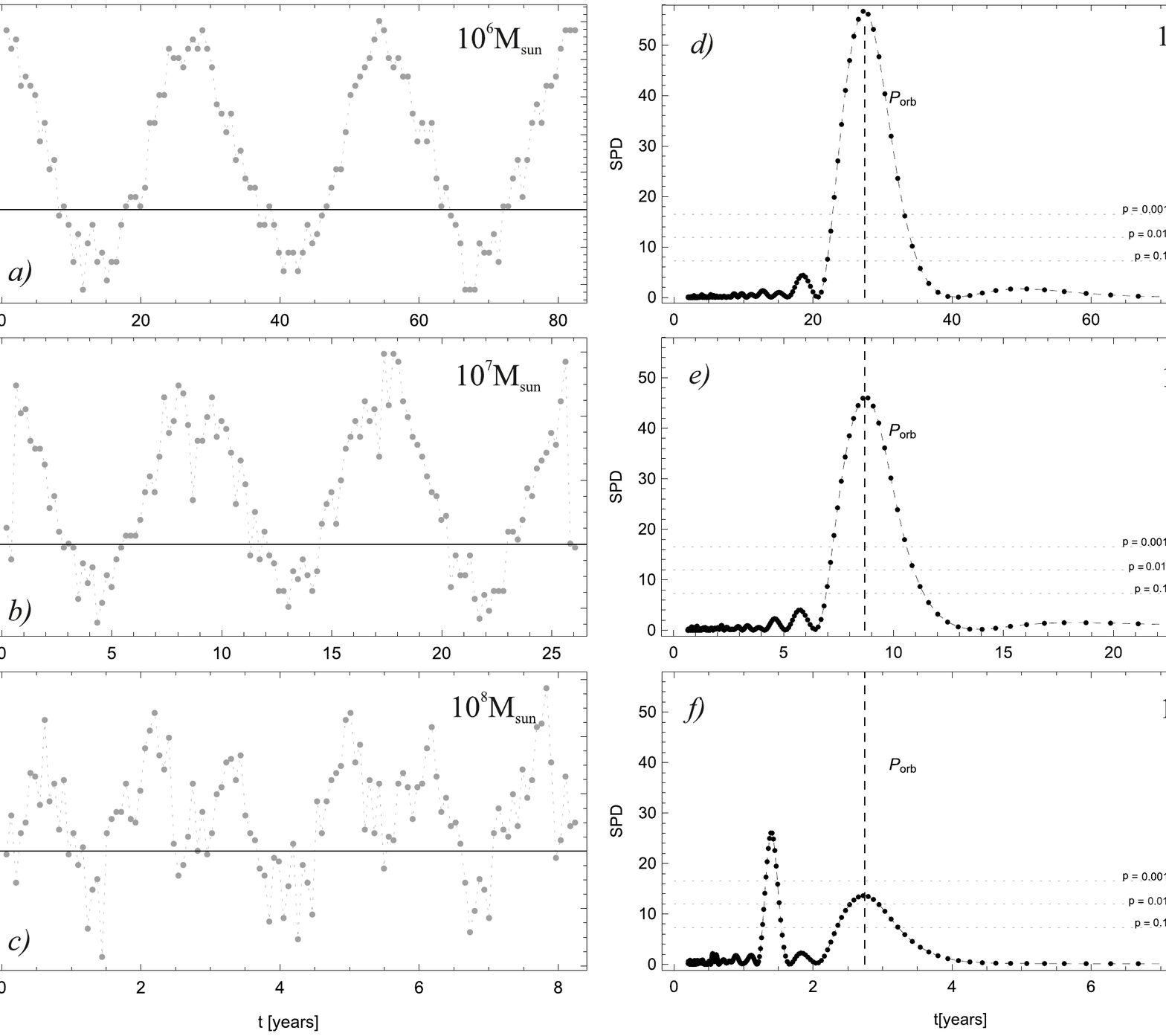}}
	\caption{The variability of the H$\beta$ line peak shift (left panels) and corresponding LS periodograms (right panels) during three full rotations of the binary system, for different total SMBBH system masses indicated in upper right corner of each plot. The horizontal significance level (dashed horizontal lines) are given for three different values of parameter p (False Discovery Rate), p = 0.1, p = 0.01, and p = 0.001. With higher Spectral Power Density (SPD) values of the peak, false discovery rate decreases. Height of the peak reaching the value of p = 0.1, indicates that there is a 10 \% probability of mistake. For lower values of SPD, probability of false discovery is bigger, and for higher SPD, false alarm has less probability.}
	\label{fig:shift}
\end{figure*}

The {\it rms} profile reveals significant variability in the line profile, which is different for different total mass of the system. In relative units, we can see that the variability is stronger in low mass systems. On the other hand, for larger total mass of SMBBH systems, changes across the line profile seem to be more prominent.

To study the line behaviour in these low mass-ratio SMBBH systems we measure the line peak shift and asymmetry for three full orbits, for three different total masses of SMBBH ($10^6, 10^7$ and $10^8\ \mathrm{M_{\odot}}$). Line asymmetry is computed as the ratio $\lambda_B/\lambda_R$ of deviations from H$\beta$ line center toward the blue $\lambda_B$ and red $\lambda_R$ line wing, at 50\% and 25\% of the peak value.

\begin{figure*}[h]
	\centerline{\includegraphics[width=16.4cm]{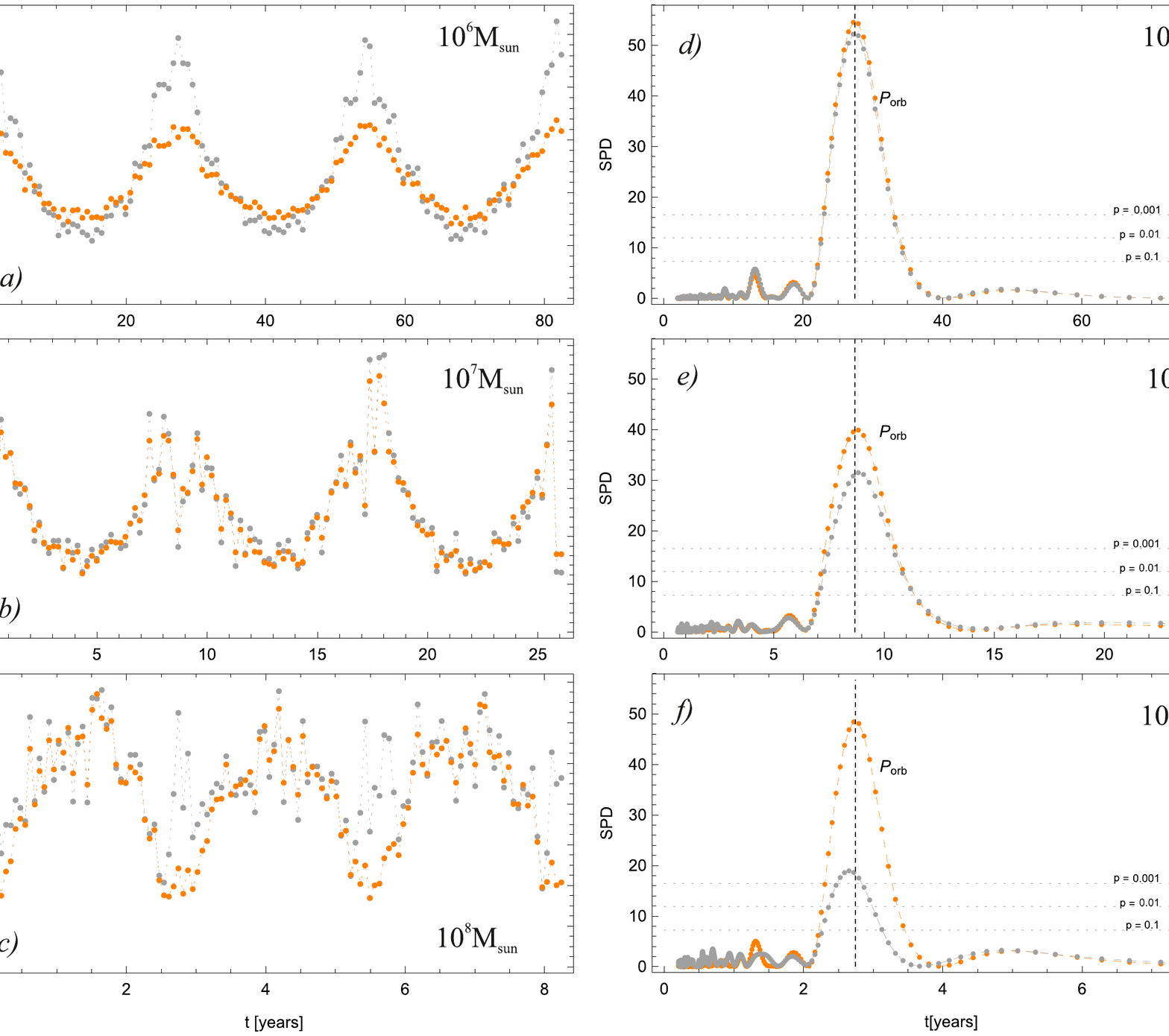}}
	\caption{Same as in Fig. \ref{fig:shift}, but for H$\beta$ line asymmetry. The line asymmetry at 50\% (gray) and 25\% (orange) of the line maximum is shown on left panels, and corresponding periodograms are given in right panels.}
	\label{fig:asym}
\end{figure*}

As stated before, the H$\beta$ line is in fact a composite line, containing contributions from the BLR of the secondary component and cBLR, therefore its behaviour is rather irregular, however periodical changes can be observed. Fig. \ref{fig:shift} shows that for the total masses of the order of $10^6\mathrm{M_{\odot}}$, the periodicity in the line-peak shift is clearly visible, whereas in the case of higher mass ($10^8\mathrm{M_{\odot}}$) the periodical variability is less pronounced due to the effect of the white noise on the line peak estimates. This effect can be seen in the computed Lomb-Scargle (LS) periodograms given in Fig. \ref{fig:shift}, right panels. The spectral power distribution (SPD) rapidly decreases for high mass binaries.

Similar behaviour is seen for the  H$\beta$ line asymmetry computed at 50\% (gray) and 25\% (orange) of the line maximum (Fig. \ref{fig:asym}). It is noticeable that the higher total SMBBH system masses produce more irregular signal with more noise.
This peculiar behaviour is due to the increase of the line widths with the increase of the SMBH masses.
With the increase of the mass, the cBLR produce stationary, but broad line, that is blended with the contribution from the BLR2, consequently giving nearly symmetric H$\beta$ line profile with very small Doppler shift. Therefore, the asymmetry and shift changes in this case is not intensive as in the case of SMBBH systems with smaller masses.
The corresponding peridograms in Fig. \ref{fig:asym}  (right panels) confirm the decrease in the periodicity for the case of the line asymmetry at the 50\% of the line maximum. However, in the case of asymmetry at 25\% of the line maximum, the periodicity can be determined for all masses with significant accuracy. This is due to the fact that the line profile is narrower at  the level of 25\% of the maximum then at the 50\%, which allow for more precise measurements.

We also simulated the variability of light curves for continuum at 5100\AA \, and H$\beta$ line. Additionally to the white noise, the intrinsic stochastic luminosity variability of each component can affect the periodicity signal of an SMBBH. To simulate this effect, the so called red noise is superposed to the generated signal. The red noise is modelled using the approach discussed in \cite{kov21} or the paper describing the PoSKI model. We added the red noise as a 30\% of light curve maximum, although different values can be adopted.
\begin{figure*}[h]
	\centerline{\includegraphics[width=16.4cm]{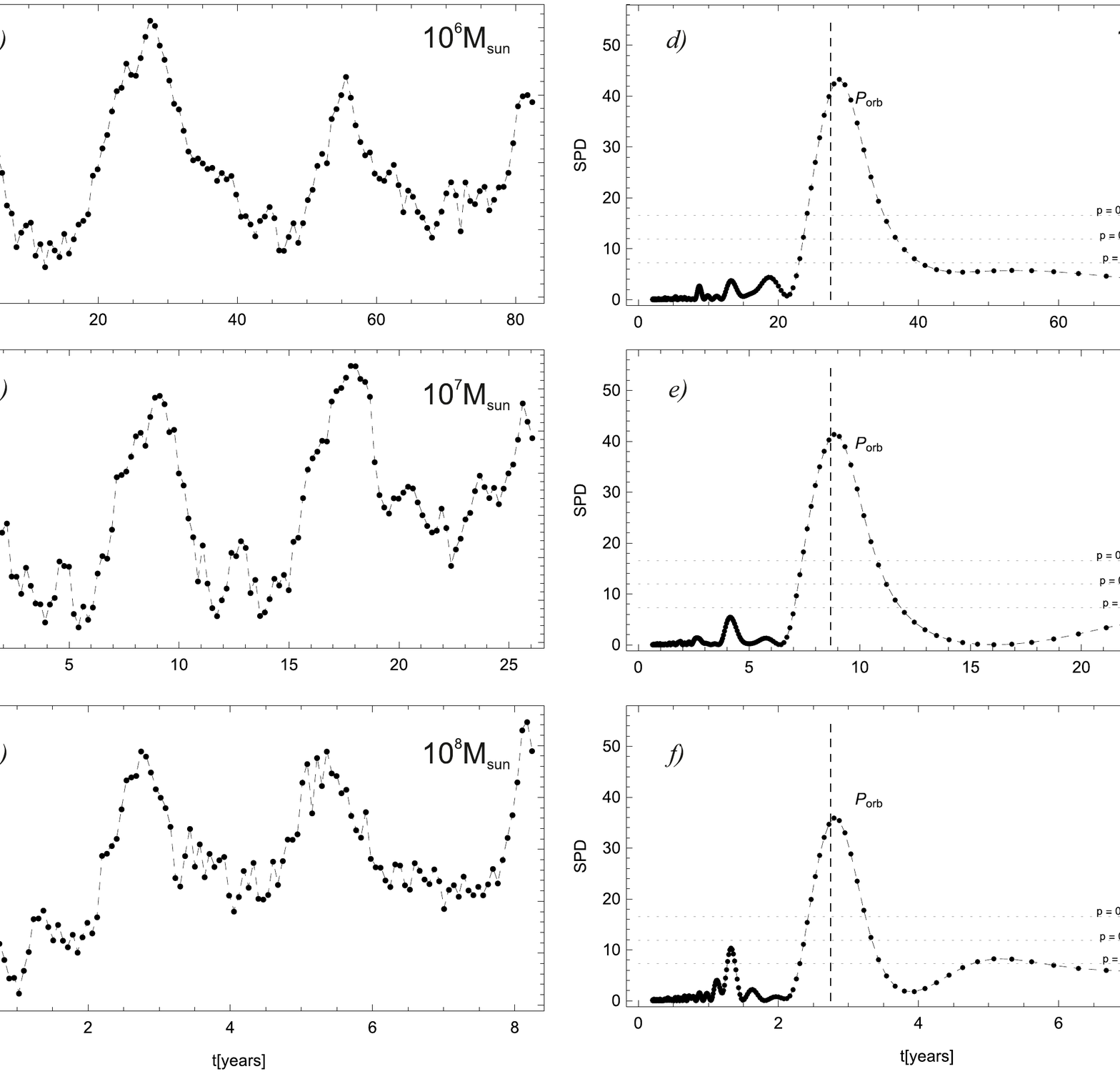}}
	\caption{Same as in Fig. \ref{fig:shift} but for the continuum at $\lambda=5100$\AA\ flux, in which both red and white noise are included. }
	\label{fig:Fvar_per}
\end{figure*}

\begin{figure*}[h]
	\centerline{\includegraphics[width=16.4cm]{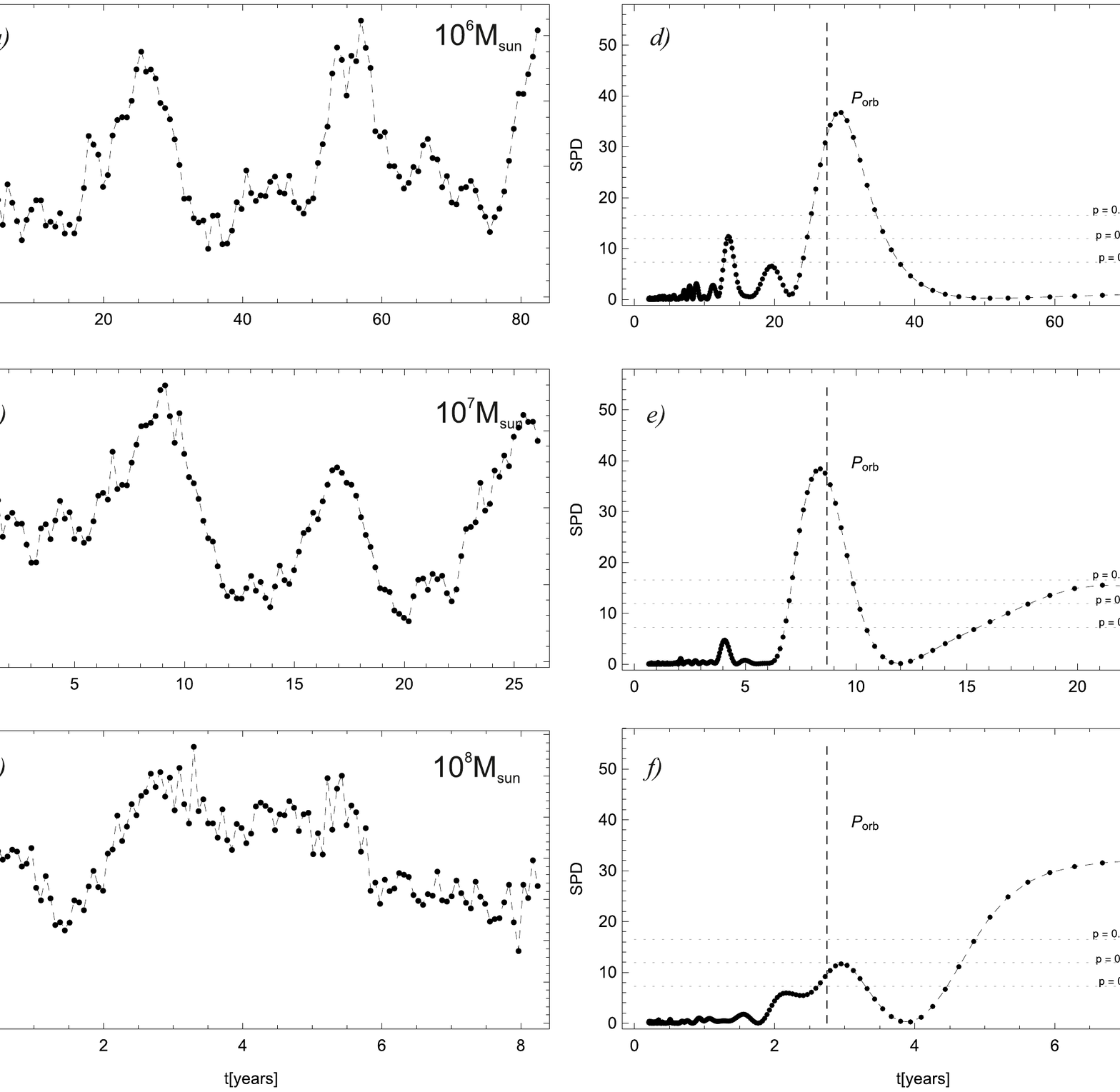}}
	\caption{Same as in Fig. \ref{fig:shift}, but for H$\beta$ line flux.}
	\label{fig:Ivar_per}
\end{figure*}

For the continuum and H$\beta$ line light curves, the periodicity is much more expressed. Figs. \ref{fig:Fvar_per} and \ref{fig:Ivar_per} show that in almost all cases we can clearly observe the periodical light curve variations. However, in case of H$\beta$ line for massive systems, the periodical trend cannot be observed, as mutual interaction between components become dominant, degrading the periodical signal, which in superposition with white and red noise becomes clearly  non-periodic.

\begin{figure}
	\centerline{\includegraphics[width=8.2cm]{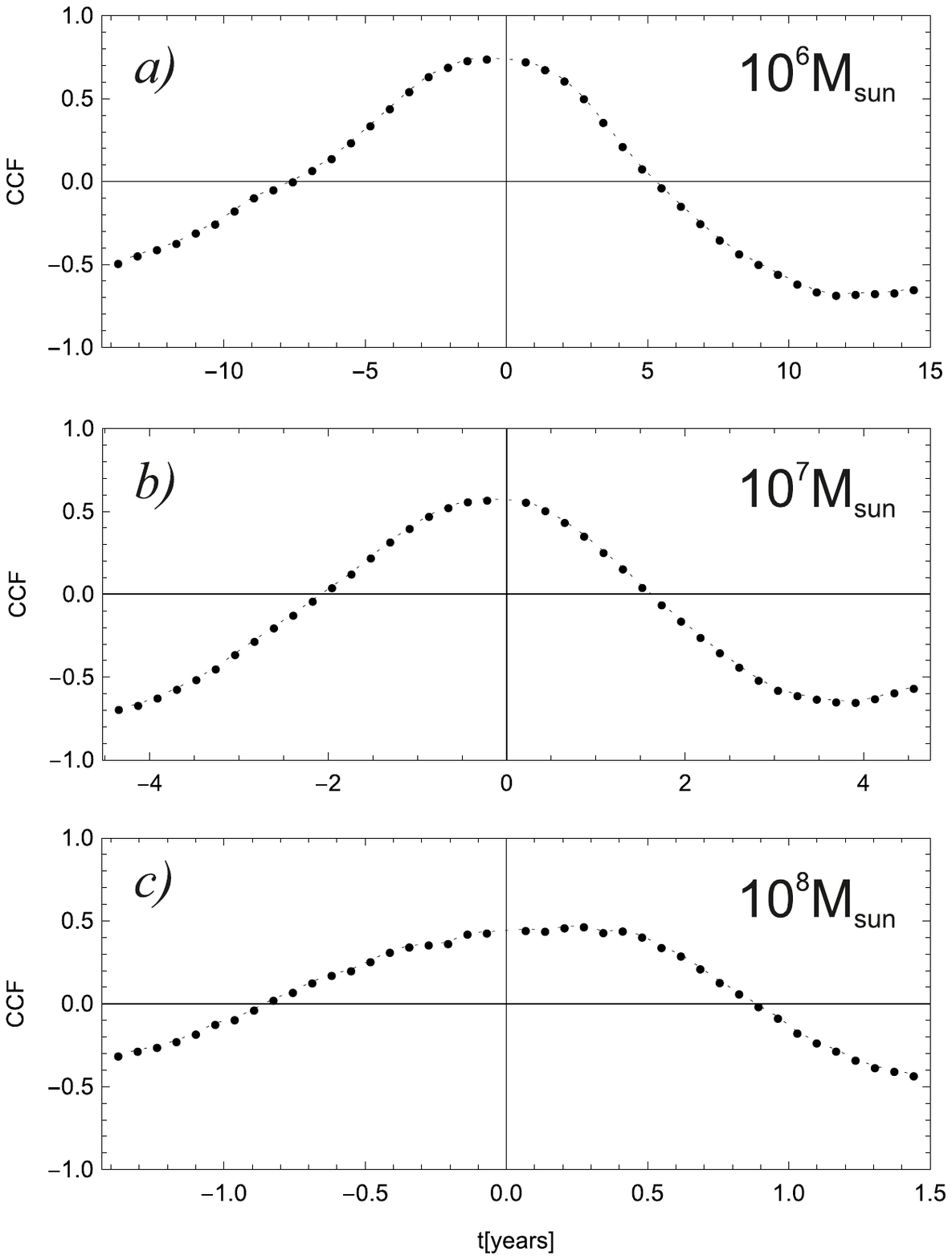}}
	\caption{Cross-correlation function between the continuum and H$\beta$ line light curves for three different total masses of binary system $10^{6}, 10^{7} \mathrm{and} 10^{8} \mathrm{M_{\odot}}$.}
	\label{fig:ccfIF}
\end{figure}
Finally, the cross correlation function for three different total SMBBH masses $10^{6}, 10^{7}\  \mathrm{and}\ 10^{8} \mathrm{M_{\odot}}$ are calculated. Figure \ref{fig:ccfIF} shows that correlation between the line and continuum is straightforward, but in the case of higher total masses, there is a noticeable time delay. For this particular case the BLR region is the largest, and thus its response to the continuum variability is delayed. Similar delay could be computed in other two cases, but on much lower time scale, and when adding more points in the calculations, i.e. increasing the light curve sampling. Additional uncertainty is invoked due to the red noise and mutual component interaction.

We note that in all computational cases we used fixed number of equally spaced calculation points, which is not the case in regular observation program. With  data reduction of real observation the possibility of period determination degrades even more, but this effect is out of the scope of this paper.

\section{Discussion}

The model which has been considered here is the special case of the PoSKI model which assumes that the one (more massive) black hole in a SMBBH system has minimal activity with low accretion ratio and without the surrounding BLR. This assumption compared with the PoSKI model gives several differences. In the PoSKI model we considered three different regions to contribute to the composite broad line emitted from an SMBBH system, but here we consider just two of them, cBLR and BLR of smaller component (BLR2). We also assumed that the super-Eddington accretion is present in the accretion disc of the smaller SMBH, while primary component is in very low accretion regime.
This causes that the broad line and continuum variability could be significant. When the total mass of the SMBBH system is $\sim 10^6\mathrm{M_{\odot}}$,
this effect is more prominent due to the narrower broad line generated by the BLR of the secondary component than by the cBLR. In this case, just the emission for the BLR of the secondary suffers Doppler shift, whereas the component from the cBLR is  stationary. As a consequence, the spectra show a small peak moving over wider stationary line component, what is easily seen in Fig. \ref{fig:composite}. This effect of the moving bump in broad line profiles is observed in some AGNs, for example in the case of NGC 4151 \citep{bon12} and also in NGC5548 \citep{shap04}.

In the cases of higher total mass of SMBBHs, the line emitted from the smaller, secondary component becomes broader, and has a weaker contribution to the total line shift and asymmetry (see Figs. \ref{fig:shift} and \ref{fig:asym}).
Depending from the phase and orientation of the orbits of SMBBHs, the moving smaller component can produce an effect in the line profile which is similar to the recoiling SMBH, where asymmetry and shift (with respect to the narrow lines) is expected \citep[as e.g. E1821+643, see ][]{robinson10,shapovalova16,jadhav21}.
Therefore, in more massive SMBBHs with observed asymmetry and shift, the SMBBH system with large differences in component masses can be an alternative explanation to the kicked-off SMBH after the coalescence phase.

The continuum light curve shows prominent periodicity in all three cases of total masses (see Fig. \ref{fig:Fvar_per}), however, the broad line flux variability, in the case of higher total masses, indicate that the periodicity can be hidden due to the red noise (see Fig. \ref{fig:Ivar_per}).
In case of higher masses, the orbital velocities becomes much higher, and therefore the influence of the red noise could be more significant in total  light curve. This effect could lead to the occasional decrease in the significance of detected periodicity for this particular case.

\section{Conclusions}
\label{sec:Conclusion}

Here we consider the model of SMBBHs with the large component mass ratio ($q\sim 0.1$) for different total masses of the system, ranging from $10^6M_{\odot}$ to $10^8M_{\odot}$, and calculated  the  spectra in the H$\beta$ wavelength band. We consider the special scenario in which only the secondary has the BLR, and there is the additionally cBLR which is ionized by accretion discs of both components and virialized by the total mass of the SMBBH system. We measured broad line components and the continuum intensity at $\lambda$5100\AA\ during three full orbits of the system. Exploring the variability of measured parameters we reach following conclusions:

1.  The modeled line profiles have large variability, showing variable asymmetry and shift. This line profile variability is more prominent in the systems with lower total mass of the system.

2. The line asymmetry and shift, as well as light curves in the line and continuum show periodicity that is caused by the dynamical period of the system.

3. The moving bump in the line profile can be detected and this bump is caused by the dynamics of the secondary, smaller component.

Finally, let us mention that in a number of AGN with broad emission lines, the bumps can be detected which may indicate a presence of a low mass ratio SMBBH system.
Additionally, large shift and asymmetry in the broad line profile also can be caused by the low mass SMBBHs as an alternative explanation of the recoiling SMBH after the merger.


\section*{Acknowledgments}

The authors acknowledge funding provided by the \fundingAgency{Astronomical Observatory Belgrade} (the contract \fundingNumber{451-03-9/2021-14/200002}), \fundingAgency{University
of Belgrade - Faculty of Mathematics} (the contract \fundingNumber{451-03-9/2021-14/200104}), \fundingAgency{University of Kragujevac - Faculty of Sciences} (the contract \fundingNumber{451-03-9/2021-14/200122}) through the grants by the Ministry of Education, Science, and Technological Development of the Republic of Serbia. D.I. acknowledges the support of the Alexander von Humboldt Foundation.We thank to the referee for very useful comments. A. K. and L. C. P. acknowledge the support by Chinese Academy of Sciences President’s International Fellowship Initiative (PIFI) for visiting scientist.
We thank the referee for very useful comments.




\bibliography{reference}%

\section*{Author Biography}

\begin{biography}
{\includegraphics[width=60pt,height=70pt]{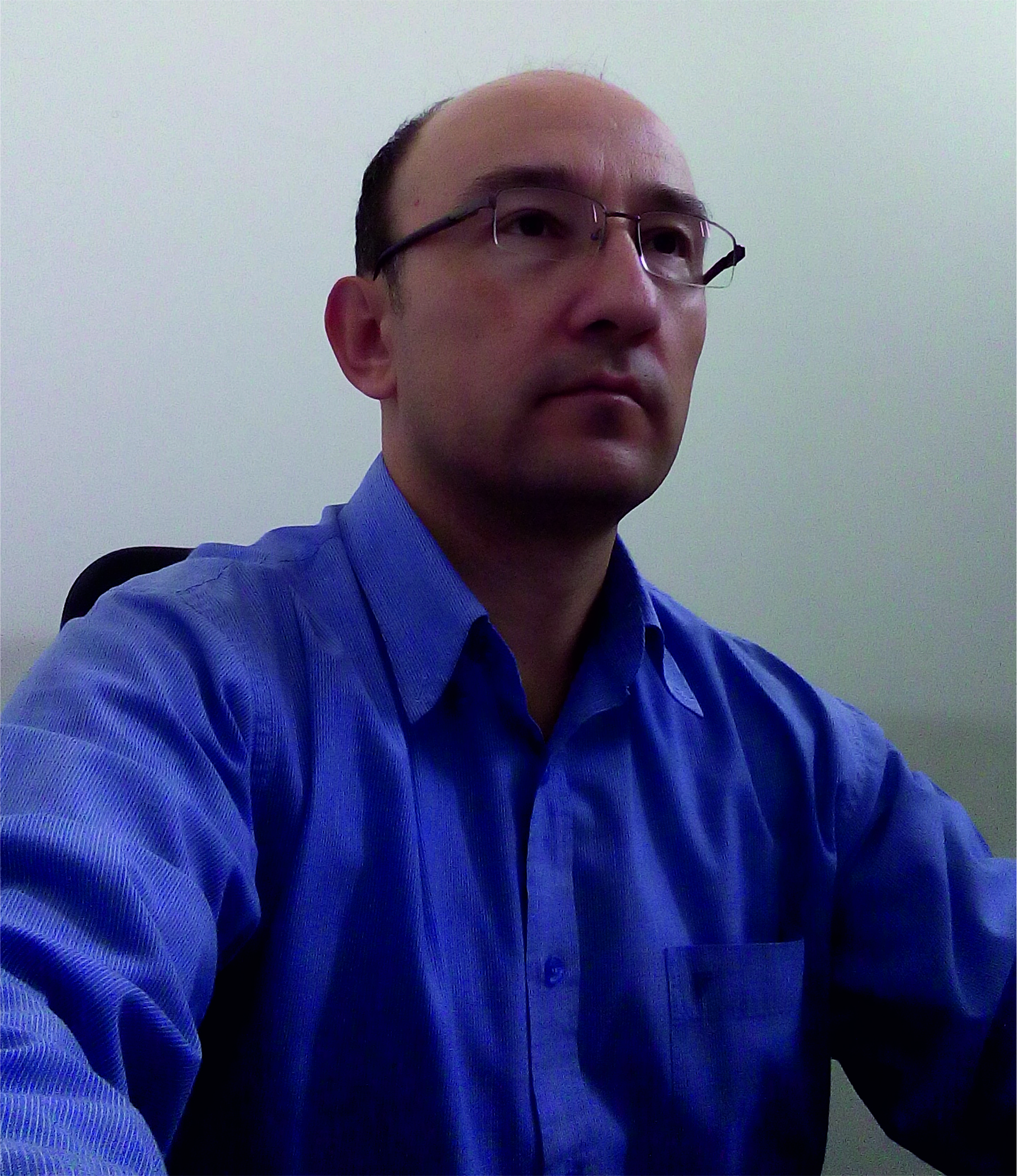}}
{\textbf{Sa{\v s}a Simi\'c} is professor at the Department of physics on the Faculty of Sciences at the University of Kragujevac. He get his PhD degree at the same University in the field of astrophysics in 2008. The area of expertise covers extra galactic and high energy astrophysics. }
\end{biography}


\end{document}